\documentclass[final,5p,times,twocolumn]{elsarticle}
\usepackage{color}
\usepackage{lipsum}
\usepackage{graphics}
\usepackage{amssymb}
\usepackage{amsmath}   % for centering equations
\journal{Physics Letters B}

\begin{document}

\begin{frontmatter}

%% Title, authors and addresses

%% use the tnoteref command within \title for footnotes;
%% use the tnotetext command for the associated footnote;
%% use the fnref command within \author or \address for footnotes;
%% use the fntext command for the associated footnote;
%% use the corref command within \author for corresponding author footnotes;
%% use the cortext command for the associated footnote;
%% use the ead command for the email address,
%% and the form \ead[url] for the home page:
%%
%% \title{Title\tnoteref{label1}}
%% \tnotetext[label1]{}
%% \author{Name\corref{cor1}\fnref{label2}}
%% \ead{email address}
%% \ead[url]{home page}
%% \fntext[label2]{}
%% \cortext[cor1]{}
%% \address{Address\fnref{label3}}
%% \fntext[label3]{}

\title{Measurement of double polarisation asymmetries in
  $\omega$-photoproduction}

\author[label:PIBN]{H.~Eberhardt}
\author[label:PIBN]{T.~C.~Jude}
\author[label:PIBN]{H.~Schmieden}
\author[label:HISKPBN,label:PNPI]{A.V.~Anisovich}
\author[label:PIBN]{B.~Bantes}
\author[label:HISKPBN,label:PNPI]{D.~Bayadilov}
\author[label:HISKPBN]{R.~Beck}
\author[label:PNPI]{Yu.~Beloglazov}
\author[label:IEBO]{M.~Bichow}
\author[label:HISKPBN]{S.~B\"ose}
\author[label:PIGI]{K.-Th.~Brinkmann}
\author[label:IPBA]{Th.~Challand}
\author[label:DPTA]{V.~Crede}
\author[label:PIGI]{F.~Diez}
\author[label:PIGI]{P.~Drexler}
\author[label:PIBN]{H.~Dutz}
\author[label:PIBN]{D.~Elsner}
\author[label:PIBN]{R.~Ewald}
\author[label:PIBN]{K.~Fornet-Ponse}
\author[label:PIGI]{St.~Friedrich}
\author[label:PIBN]{F.~Frommberger}
\author[label:HISKPBN]{Ch.~Funke}
\author[label:HISKPBN]{M.~Gottschall}
\author[label:PNPI]{A.~Gridnev}
\author[label:HISKPBN]{M.~Gr\"uner}
\author[label:HISKPBN,label:PIGI]{E.~Gutz}
\author[label:HISKPBN]{Ch.~Hammann}
\author[label:PIBN]{J.~Hannappel}
\author[label:HISKPBN]{J.~Hartmann}
\author[label:PIBN]{W.~Hillert}
\author[label:HISKPBN]{Ph.~Hoffmeister}
\author[label:HISKPBN]{Ch.~Honisch}
\author[label:IPBA]{I.~Jaegle}
\author[label:HISKPBN]{D.~Kaiser}
\author[label:HISKPBN]{H.~Kalinowsky}
\author[label:HISKPBN]{F.~Kalischewski}
\author[label:PIBN]{S.~Kammer}
\author[label:IPBA]{I.~Keshelashvili}
\author[label:PIBN]{V.~Kleber}
\author[label:PIBN]{F.~Klein}
\author[label:HISKPBN]{E.~Klempt}
\author[label:HISKPBN]{K.~Koop}
\author[label:IPBA]{B.~Krusche}
\author[label:HISKPBN]{M.~Kube}
\author[label:HISKPBN]{M.~Lang}
\author[label:PNPI]{I.~Lopatin}
\author[label:IPBA]{Y.~Maghrbi}
\author[label:PIGI]{K.~Makonyi}
\author[label:PIGI]{V.~Metag}
\author[label:IEBO]{W.~Meyer}
\author[label:HISKPBN]{J.~M\"uller}
\author[label:PIGI]{M.~Nanova}
\author[label:HISKPBN,label:PNPI]{V.~Nikonov}
\author[label:PIGI]{R.~Novotny}
\author[label:HISKPBN]{D.~Piontek}
\author[label:PIBN]{S.~Reeve}
\author[label:IEBO]{G.~Reicherz}
\author[label:IPBA]{T.~Rostomyan}
\author[label:PIBN]{S.~Runkel}
\author[label:HISKPBN,label:PNPI]{A.~Sarantsev}
\author[label:HISKPBN]{St. Schaepe}
\author[label:HISKPBN]{Ch.~Schmidt}
\author[label:HISKPBN]{R.~Schmitz}
\author[label:HISKPBN]{T.~Seifen}
\author[label:HISKPBN]{V.~Sokhoyan}
\author[label:PNPI]{V.~Sumachev}
\author[label:HISKPBN]{A.~Thiel}
\author[label:HISKPBN]{U.~Thoma}
\author[label:HISKPBN]{M.~Urban}
\author[label:HISKPBN]{H.~van Pee}
\author[label:HISKPBN]{D.~Walther}
\author[label:HISKPBN]{Ch.~Wendel}
\author[label:IEBO]{U.~Wiedner}
\author[label:HISKPBN]{A.~Wilson}
\author[label:HISKPBN]{A.~Winnebeck}
%(The CBELSA/TAPS collaboration)
%\author{\\(The CBELSA/TAPS collaboration)}

\address[label:PIBN]{Physikalisches Institut, Universit\"at Bonn, Germany}
\address[label:HISKPBN]{Helmholtz-Institut f\"ur Strahlen- und Kernphysik,
  Universit\"at Bonn, Germany}
%\address[label:PNPI]{Petersburg Nuclear Physics Institute, Gatchina, Russia}
\address[label:PNPI]{NRC Kurchatov Institute, Petersburg Nuclear Physics Institute, Gatchina, Russia}
\address[label:IEBO]{Institut f\"ur Experimentalphysik I, Ruhr--Universit\"at
  Bochum, Germany}
\address[label:PIGI]{II. Physikalisches Institut, Universit\"at Gie{\ss}en,
  Germany}
\address[label:IPBA]{Institut f\"ur Physik, Universit\"at Basel, Switzerland}
\address[label:DPTA]{Department of Physics, Florida State University,
  Tallahassee, USA}

\begin{abstract}
%% Text of abstract
The first measurements of the beam-target-helicity-asymmetries \textit{E} and \textit{G} in the 
photoproduction of $\omega$-mesons off protons at the CBELSA/TAPS experiment are reported. \textit{E} (\textit{G}) was measured using circularly (linearly) polarised photons and a longitudinally polarised target.  \textit{E} was measured over the photon energy range from close to threshold ($E_\gamma = 1108$~MeV) to $E_\gamma = 2300$~MeV and \textit{G} at a single energy interval of $1108 < E_\gamma <1300$~MeV. Both measurements cover the full solid angle. The observables \textit{E} and \textit{G} are highly sensitive to the contribution of baryon resonances, with \textit{E} acting as a helicity filter in the $s$-channel. The new results indicate significant $s$-channel resonance contributions together with contributions from $t$-channel exchange processes. A partial wave analysis reveals strong contributions from the partial waves with spin-parity $J^P=3/2^+, 5/2^+$, and $3/2^-$.
\end{abstract}

\begin{keyword}
%% keywords here, in the form: keyword \sep keyword
Meson production \sep Polarisation in interactions and scattering \sep Light
mesons (S=C=B=0)
%% MSC codes here, in the form: \MSC code \sep code
%% or \MSC[2008] code \sep code (2000 is the default)

\end{keyword}

\end{frontmatter}

%%
%% Start line numbering here if you want
%%
% \linenumbers

%% main text
\section{Introduction}
\label{labelIntro}
The excitation spectrum of the nucleon has long been studied to understand the non-perturbative regime of QCD, however this still remains poorly understood. In particular, constituent quark models \cite{IK,CI,LKMP} predict significantly more states than experimentally observed \cite{PDG}. This is sometimes referred to as the ``missing resonance problem'' and is most noticeable for relatively high lying states. However, masses and parity orderings of some low lying states are also not well reproduced. These deficits also appear in present Lattice-QCD calculations and may be due to the fact that the models being used are not fully implementing a treatment of resonance decay \cite{Edwards:2011jj}.  The inclusion of resonance decays via meson-baryon couplings may affect both the number and ordering of the states 
\cite{Doring:2011ip,Lang:2012db}.

The missing resonance problem may also be related to experimental shortcomings. By far most of the observed states have been discovered in pion induced processes and therefore states with small $\pi N$ couplings may have escaped detection \cite{CR}. The photoproduction of mesons, in particular non-pionic final states, may therefore provide a tool to investigate the existence of hitherto unobserved resonances.

The photoproduction of $\omega$ mesons is suitable to address this 
issue because the reaction threshold lies in the lesser explored third
resonance region.
Furthermore, the $\omega$ is isoscalar ($I = 0$).
Therefore, in $s$-channel processes, only $N^*$ resonances ($I = \frac{1}{2}$) couple to the nucleon ground state, with no interference from
$\Delta^*$ states ($I = \frac{3}{2}$).
This greatly simplifies the complexity of the
contributing excitation spectrum.

Due to the vector character of the
$\omega$ meson, at least 23 independent observables
have to be measured to achieve a complete set of observables with respect to the
decomposition of the reaction amplitudes \cite{SAR}.
This is much more involved than in pseudoscalar meson photoproduction
where, in principle, only 8 observables suffice, however it is similar to other
channels such as double pseudoscalar meson photoproduction.
It is well known that $t$-channel processes dominate $\omega$
photoproduction at high energies.
However, in the threshold vicinity, previous experiments indicate that $s$-channel
processes also contribute (see for example Refs.~\cite{AJ,FK, Dietz15}).
Individual double polarisation observables may act as
sensitive probes to disentangle these processes, even if a complete set of observables
is not yet available \cite{SAR}.

A comprehensive study of $\omega$ photoproduction using an
unpolarised liquid hydrogen target and the ``charged'' \mbox{$\omega \rightarrow % \mbox keeps the decay on the same line
\pi^+\pi^-\pi^0$} decay was performed at  CLAS \cite{WI1,WI2}.
Evidence for contributions from $s$-channel resonances $N(1680)5/2^+$ and $N(1700)3/2^-$ was found near threshold, and contributions from $N(2190)7/2^-$ were strongly supported.
The data also supported $5/2^+$ resonance states around 1.9-2.0~GeV and $3/2^+$ states around 1.8-2.0~GeV.
The goal of the present investigation was to
further study  the possible role of $s$-channel excitations in the threshold
region through the measurement of double polarisation observables.
The Bonn Frozen Spin hydrogen (butanol) target \cite{BD,Dutz:2004eb}  was used
in longitudinal polarisation mode,
in combination with linearly and circularly polarised
photon beams.
The experiments were performed at the ELSA electron accelerator \cite{Hillert06}
at the Physics Institute of Bonn University.
Using the CBELSA/TAPS detector setup, the ``neutral'' decay
$\omega \rightarrow \pi^0\gamma$ was identified, which
ideally suits the detector capabilities.%\textcolor{red}{ and is complementary to the charged decay \cite{ZAC}.}

The paper is organised as follows.
Sec.~\ref{labelMech} discusses the double polarisation observables relevant
to this study.
The experiment is briefly described in Sec.~\ref{labelExp}
and the data analysis in Sec.~\ref{labelAna},
before the results are presented in Sec.~\ref{labelRes}.
The paper concludes with a summary and outlook in Sec.~\ref{labelSA}.

\section{Double Polarisation Observables and the Mechanism of $\omega$ photoproduction}
\label{labelMech}

\begin{figure}[]
  \begin{center}
    \includegraphics[width=\columnwidth]{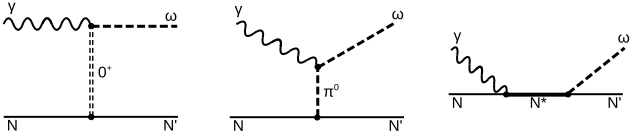}
  \end{center}
  \caption{\footnotesize %
    $\omega$ production via $t$-channel $0^{+}$ (Pomeron) exchange
    (left), $t$-channel $\pi^{0}$ exchange (middle) and $s$-channel intermediate
    resonance (right).
    }
  \label{fig:ProdMech}
\end{figure}

It is mandatory to understand
the reaction dynamics in order to extract resonance information from
$\omega$ photoproduction.  At high photon energies, $\omega$ production is dominated by
diffractive scattering.
The fluctuation of the incoming photon into a $q \bar q$-pair
produces the vector meson in the vicinity of a strongly interacting
recoil partner, mediated through the exchange of natural
parity quantum numbers of the \textit{Pomeron}
(Fig.~\ref{fig:ProdMech} (left)).
The cross section shows a characteristic
exponential fall off with squared recoil momentum, \textit{t}.  Significant unnatural parity $\pi^{0}$-exchange
(Fig.~\ref{fig:ProdMech} (middle)) has been expected due to the sizeable
$\omega\rightarrow\pi^{0}\gamma$ decay (8.3~\% branching ratio) and was indeed reported \cite{WI2,BA}.
Meson exchange models of $\omega$ photoproduction \cite{FS} have predicted dominant pion exchange processes near threshold (for photon beam energies less than 2~GeV),
however a recent partial wave analysis finds a negligible contribution (see below).
Neither Pomeron
nor $\pi^{0}$-exchange however, are able to reproduce the strong threshold energy
dependence of the cross section and the observed $\omega$ decay angular
distribution (see for example Ref.~\cite{WI2,BAR}).
This may suggest $s$-channel contributions (Fig.~\ref{fig:ProdMech} (right)), which is further corroborated by measurements of the photon beam asymmetry,
$\Sigma$ \cite{AJ,FK}.

\subsection{Circularly Polarised Beam}
\label{sec:CircBeamObservable}

For the combination of circularly polarised beam and longitudinally polarised nucleon
target, the cross section can be written in the form

\begin{equation} \label{eq:crosssection}
  \frac{d\sigma}{d\Omega} = \frac{d\sigma_{0}}{d\Omega}
  (1 - P^{\odot}_{\gamma}P^{z}_{T}E)\,.
\end{equation}

$\sigma_{0}$ denotes the unpolarised cross section,
$P^{\odot}_{\gamma}$ the degree of circular beam polarisation,
and $P^{z}_{T}$ the degree of longitudinal target polarisation.
$E$ is the beam-target helicity asymmetry.
The sensitivity of $E$ to the reaction mechanism
is shown in Ref.~\cite{HS} in an intuitive way:
For vector meson photoproduction, it is important which hadron couples to the
polarised photon.
In the case of Pomeron or $\pi^{0}$-exchange
(Fig.~\ref{fig:ProdMech} left and middle), the photon couples to the vector
meson directly but not to the polarised target.
With no angular momentum exchanged in the $t$-channel,
this leads to a zero beam-target asymmetry.
Conversely, in the case of $s$-channel production, the photon directly couples to the polarised nucleon.
In this case, the helicity asymmetry will reflect the projection onto the beam axis of the spin of the
intermediate $s$-channel state.
Such a behaviour is predicted in Ref.~\cite{SAR}.
In the case of mixing Pomeron and $\pi^{0}$ exchange,
$E$ may also be non-zero, with a linear dependence
in $\cos{\theta^{\omega}_{CMS}}$ \cite{SAR}.

\subsection{Linearly Polarised Beam}
\label{sec:LinBeamObservables}

Combining a linearly polarised beam and longitudinally polarised
target, using the notation of Ref.~\cite{SAR}, the two beam-target asymmetries
$G$ and $G_\pi$ can be extracted. $G$ is the target asymmetry associated with the azimuthal asymmetry of the produced $\omega$-meson, and $G_\pi$ with that of the $\pi^0$ of the neutral decay.  

Previous data for $\omega$ photoproduction at the CBELSA/TAPS-experiment were taken using an unpolarised target.  Spin density matrix elements were extracted from this data and the results are described in Ref.~\cite{Wilson:2015tbd}.

\section{CBELSA/TAPS-experiment}
\label{labelExp}

Electrons from ELSA with an energy ($E_0$)
of 2.4 and 3.2~GeV (for circular or linear polarisation respectively) were used to produce photons via
bremsstrahlung off a thin radiator.  To measure the photon energy,
electrons which radiated a photon were momentum analysed using a
magnetic dipole (tagging-) spectrometer,
covering a photon energy range of
$E_{\gamma} = (0.175-0.98)E_{0}$ \cite{KFP}.

Longitudinally polarised electrons were used to produce circularly polarised photons.
A M\o ller polarimeter was integrated into the tagging spectrometer,
using a 20~$\mu$m thick magnetised foil which simultaneously
acted as a bremsstrahlung radiator and a M\o ller target.
Symmetric M\o ller pairs emitted perpendicular to the dispersive
plane of the tagging spectrometer were momentum selected by a pair of lead-glass detectors behind the tagger magnet.
With this setup the electron beam polarisation was
measured to between 60 - 65\% during the duration of the data taking, with a relative uncertainty of approximately 2\% \cite{KAM}.
The degree of polarisation transfer from the beam electron to the
radiated photon can then be calculated \cite{Olsen59}.  
As a guide, using an electron beam energy of 2.4~GeV, the absolute circular polarisation of the photon beam
was 40\% and 62\% at photon beam energies of 1200~MeV and 2200~MeV respectively.

A 500~$\mu$m thick diamond radiator was used to produce linearly polarised photons~\cite{Timm69}.
The radiator was aligned relative to the incident electron beam to select the plane
of polarisation and the energy of the coherent edge.  The coherent peaks were set at photon energies of
950, 1150 and 1350~MeV.  
The degree of polarisation was determined using the Analytical Bremsstrahlung Calculation (ANB) software~\cite{Natter03},
with a typical maximum degree of linear polarisation of 50\%, 
accurate to a relative
systematic error of 5\%.  Ref.~\cite{Elsner09} describes the method
of coherent bremsstrahlung and the performance of the setup.

The linearly or circularly polarised photon beam was incident upon a 2~cm long
longitudinally polarised butanol (C$_{4}$H$_{10}$O) target \cite{BD}. The
degree of target polarisation was measured via NMR-techniques and was approximately
70\% on average, with a 2\% relative systematic error.

A three layer scintillating fibre detector \cite{SUF} to identify charged
particles surrounded the target within the acceptance of the Crystal-Barrel
calorimeter \cite{AKE}. This calorimeter consisted of 1230 CsI(Tl) crystals,
cylindrically arranged around the target and covering a polar angular range of
30 to 150 degrees. The detector was complemented by a forward cone detector
of the same material, which was assembled with scintillating plates for charge
identification, covering a polar angular range of 11.2 to 27.5 degrees
\cite{FUN,WEN}.

The 1 to 12 degrees forward cone was covered by the Mini-TAPS detector, set up
in a hexagonally shaped wall of 216 BaF$_{2}$ crystal modules, also assembled with
scintillating plates for charged particle identification.

The whole setup was able to detect charged as well as neutral particles, however
it was optimised for the detection of photons. The total coverage is about
96~\% of the whole solid angle in the laboratory frame.

\section{Data analysis}
\label{labelAna}

The $\omega$ was identified through its decay to $\pi^{0}\gamma$. Thus during
offline analysis, four detector hits were required, corresponding to three
photons and the proton.  The proton (charge)
identification was done using the signals of the inner scintillating fibre
detector or the scintillating plates of the forward cone and the Mini-TAPS
detector. The reconstructed angles of the protons were used, however the energy
information from the calorimeters was disregarded, since the detector response
was very different for photons and high energy ($> 400$~MeV) protons. 

Timing cuts according to detector resolutions were applied
between the tagged incident photon beam and energy deposits in the detectors.
 The invariant mass of the summed four momenta of
two of the photons was required to be between 105-165~MeV (a 3$\sigma$ fit
due to detector resolutions to the $\pi^0$ mass).  The invariant mass of the reconstructed
$\pi^0$ and the other photon was required to be within 3$\sigma$ of the $\omega$ mass.
 There was a small amount of
background from the $\gamma p \rightarrow \pi^0 p$ channel, where a $\pi^0$
decay photon caused an extra ``split-off'' cluster due to the electromagnetic shower
in the crystal.  These events were removed from the data sample by requiring 
that the photon not originating from the $\pi^0$ decay had an energy greater
than 200~MeV.  Further kinematic cuts were applied in order to ensure
longitudinal and transverse momentum conservation. 

\begin{figure}[]
%\vspace{-20mm}
 \begin{center}
 	\includegraphics[trim=0cm 0cm 0cm 0cm,clip=true,width=1\columnwidth]{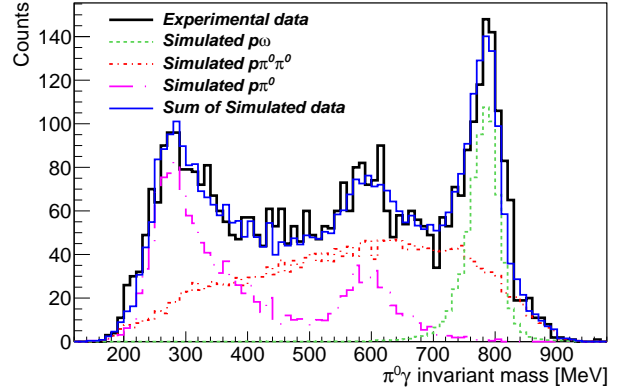}
  \end{center}
  \caption{\footnotesize %
 Typical $\pi^{0}\gamma$ invariant mass distribution of one bin
    ($E_{\gamma} = 1300 - 1400$~MeV, $\cos{\theta^{\omega}_{CMS}} = (-0.75) -
    (-0.5)$).  Experimental and simulated data labelled inset.  Colour available online.
}
  \label{fig:invmass}
\end{figure}

After all selection cuts, a $\pi^{0}\gamma$ invariant mass spectrum as shown in
Fig.~\ref{fig:invmass} was obtained. Monte Carlo simulations of
signal and background events showed that the dominating background
channels originated from $\pi^{0}$ and $2\pi^{0}$ production.   
In the $\omega$ invariant mass range however, only $2\pi^{0}$ was significant for all beam energy and polar angle bins. 
These background events also carried sizeable asymmetries,
which needed to be corrected for.  A dedicated analysis of this channel was performed to extract the asymmetry for every
kinematic bin.  The fraction of $2\pi^{0}$ background under the $\omega$ mass peak was determined by fitting Monte Carlo spectra
to the experimental data as in Fig.~\ref{fig:invmass}.  The asymmetry from the $2\pi^{0}$ background was then scaled accordingly and subtracted
to leave the asymmetry from the $\omega$ channel.

The beam-target-helicity asymmetry, $E$, was extracted by the combination of the
two different datasets, with either parallel data ($N^{\uparrow\uparrow}$), when the beam and target polarisations point in the same direction,
 or antiparallel data ($N^{\uparrow\downarrow}$), when the polarisation directions are opposite:

\begin{center}
\begin{equation}\label{eq:asymmetry}
P^{\odot}_{\gamma}P^{z}_{T}E =   \frac{N^{\uparrow\downarrow} - N^{\uparrow\uparrow}}{N^{\uparrow\uparrow} +
    N^{\uparrow\downarrow}}
\end{equation}%
\end{center}

The beam-target asymmetry using a linearly polarised beam, $G$ (and $G_\pi$ when measuring the asymmetry of the decay $\pi^0$) was determined by measuring the yield ($N$) as a function of the azimuthal angle between the meson and the target polarisation direction ($\psi$).  
This was repeated for two different azimuthal directions of beam polarisation ($\phi_{\gamma,l} = +45^0, -45^0$) and either target polarised parallel ($P^T_z$) or antiparallel ($P^T_{-z}$) to the beam direction.  $G$ was then extracted from a combined asymmetry of the four combinations:
\begin{minipage}{\linewidth} % the following should stay on the same page
\begin{equation}\label{eq:asymmetry}
%\begin{split}
 -P^l_\gamma P^{z}_{T}G\cos(2\psi) = 
 % \frac{[N(+45^0,P^T_z) + N(-45^0,P^T_{-z})] - [N(+45^0,P^T_{-z}) + N(-45^0,P^T_{z})]}{[N(+45^0,P^T_z) + N(-45^0,P^T_{-z})] + [N(+45^0,P^T_{-z}) + N(-45^0,P^T_{z})]}
%\end{split}
\end{equation}% 
\begin{equation}\label{eq:asymmetry}
%\begin{split}
 %-P^l_\gamma P^{z}_{T}G\cos(2\psi) =  \nonumber\nonumber
  \frac{[N(+45^0,P^T_z) + N(-45^0,P^T_{-z})] - [N(+45^0,P^T_{-z}) + N(-45^0,P^T_{z})]}{[N(+45^0,P^T_z) + N(-45^0,P^T_{-z})] + [N(+45^0,P^T_{-z}) + N(-45^0,P^T_{z})]}\nonumber
%\end{split}
\end{equation}%
 \end{minipage}
%\end{center}
%where $\psi$ is the angle between the target polarisation and meson, orthogonal to the beam direction.

\begin{figure}[]
%\vspace{-20mm}
 \begin{center}
    \includegraphics[trim=16cm 0cm 0cm 0cm, clip=true,width=0.8\columnwidth]{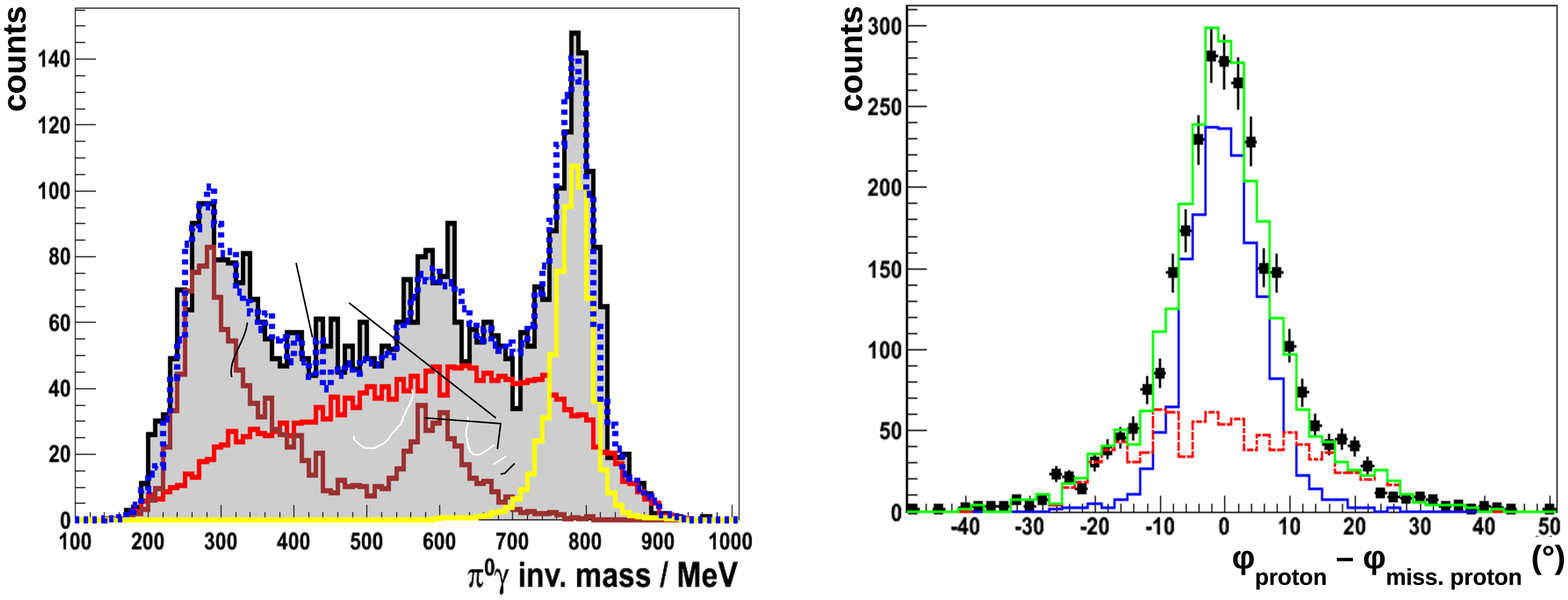}
  \end{center}
  \caption{\footnotesize %
    Spectrum used for the determination of the ``dilution factor'': The
    azimuthal angular difference of the detected proton and the calculated proton
    direction (using missing momentum techniques) is shown. The curves
    represent the liquid hydrogen data (blue or dark grey solid line), carbon data (red or dark grey dashed line) and sum of carbon
    and liquid hydrogen data (green or light grey solid line) in comparison to the butanol data (black squares). Colour available online.
    }
  \label{fig:dilfit}
\end{figure}

The polarised target provided a complication to the analysis. The frozen spin butanol (C$_4$H$_{10}$O) target \cite{Dutz:2004eb} contained the polarised hydrogen atoms in which the atomic electron polarisation was transferred dynamically to the free protons.  A mean polarisation, monitored via NMR techniques, of about 70\% was reached.  The protons bound in the carbon and oxygen nuclei however remained unpolarised.
The contribution of the bound protons (through quasifree processes) required a correction to the measured target polarisation by what is referred to herein as the ``dilution factor''. 
The effective dilution factor is related to the relative contribution of quasifree production, which strongly depends 
on the widths of the applied kinematic cuts, on the energy of the beam photon, and on the polar angle of the $\omega$.  This contribution is determined by separate measurements on carbon and hydrogen targets. These data are normalised, using the spectra described in Fig.~\ref{fig:dilfit}, so that the butanol distribution agrees with the sum of liquid hydrogen and Fermi broadened carbon distributions \cite{Thiel:2012yj}.

Approximately 225k and 5k events were used to determine $E$ and $G$ over the measured kinematic ranges respectively.  
The data was distributed towards forward angles
due to the diffractive nature of the cross section.  The statistical error per kinematic bin has contributions from the number of reconstructed $\omega$ events, and the number of subtracted
background from 2$\pi^0$ events.%, and the subtraction of the carbon data to account for the dilution factor.

The systematic errors consist of uncertainties in the background correction, polarisation determinations and the determination of the
dilution factor.  The systematic uncertainty between the relative flux of the two polarisation settings was negligible.  Furthermore, systematic effects concerning the analysis conditions by the variation of kinematic cut ranges were studied.  Individual systematic uncertainties were added linearly for a conservative estimation of the final systematic errors.

A more detailed description of the data analysis can be found in Ref.~\cite{EBE}.

\section{Results and interpretation}\label{labelRes}

\begin{figure}[]
  \begin{center}
\includegraphics[width=1\columnwidth]{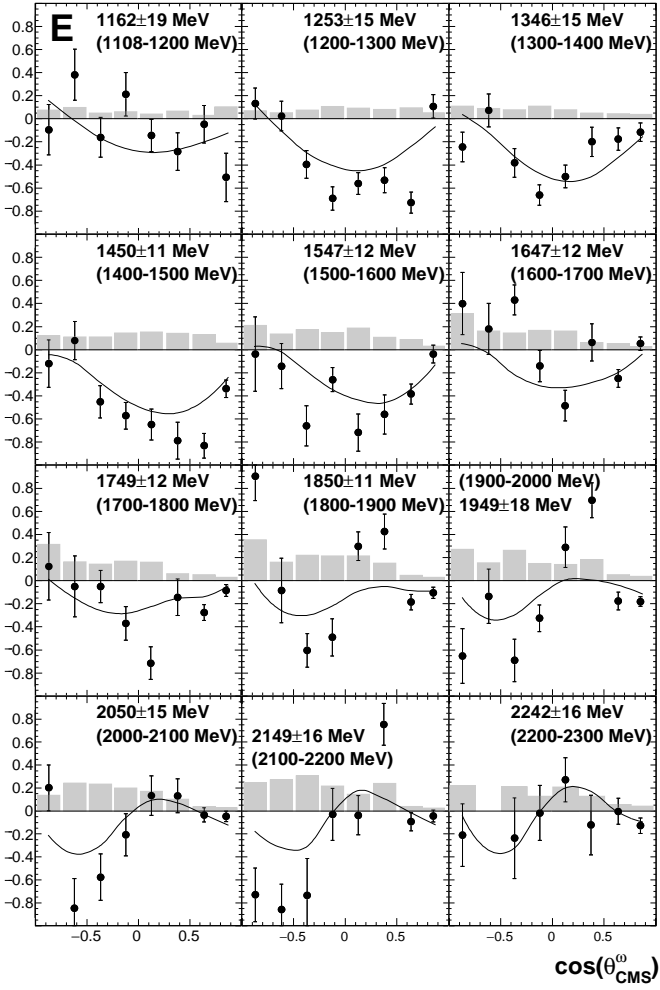}
  \end{center}
  \caption{\footnotesize %
    Beam-target-helicity asymmetry, $E$, as a function of
    $\cos{\theta^{\omega}_{CMS}}$. Systematic errors are on the abscissa.
    The event weighted average beam energy and systematic error is given for each energy interval, with the energy range given in parentheses.
    The solid line is the result of the
    Bonn-Gatchina PWA when including this data (see text for details).  The data are tabulated in Ref.~\cite{EBE}.
  }
  \label{fig:resultsEangle}
\end{figure}

\begin{figure}[]
  \begin{center}
    \includegraphics[width=1\columnwidth]{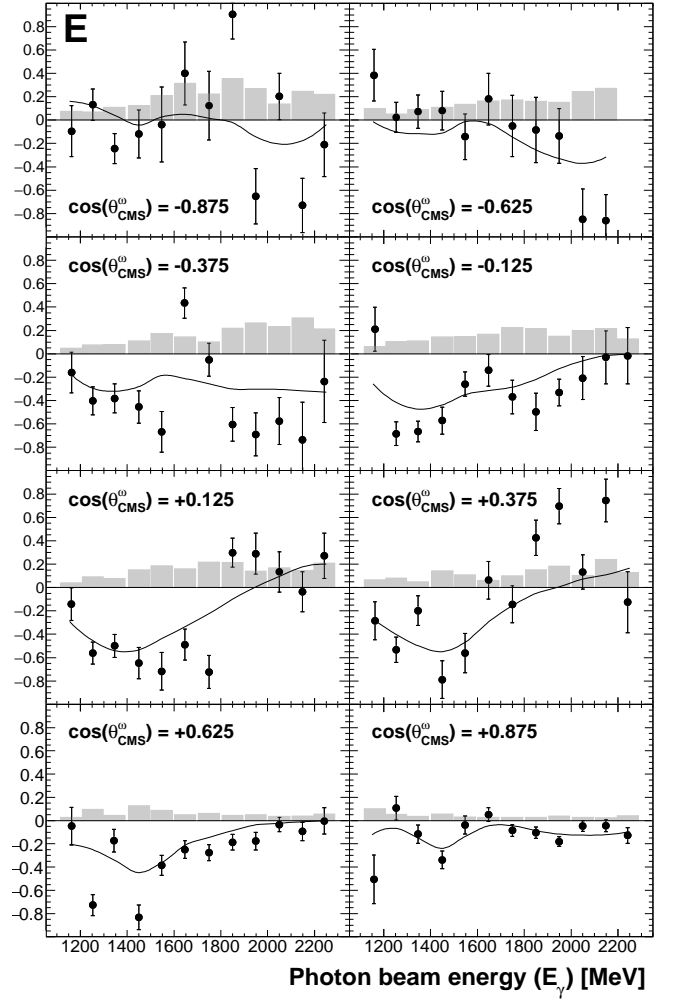}
  \end{center}
  \caption{\footnotesize %
    Beam-target-helicity asymmetry, $E$, as a function of
    photon beam energy (the same data as in Fig.~\ref{fig:resultsEangle}). Systematic errors are on the abscissa.
    The solid line is the result of the
    Bonn-Gatchina PWA when including this data (see text for details).
  }
  \label{fig:resultsEenergy}
\end{figure}

Data for the beam-target-helicity asymmetry, $E$,  are shown in Fig.~\ref{fig:resultsEangle} and \ref{fig:resultsEenergy} for
centre-of-mass-energies from 1720~MeV to 2280~MeV. At forward angles where $t$-channel exchange is expected to dominate the reaction, 
$E$ is close to zero.  This is expected for pure pion or Pomeron exchange but incompatible with mixed pion and Pomeron exchange \cite{SAR}. At more backward angles, the data show a clear nonlinear behaviour in $\cos(\theta^{\omega}_{CMS})$, indicating significant resonance
contributions to the $\omega$ production channel.

\begin{figure}[]
  \begin{center}
    %\hspace{-3mm}\includegraphics[width=1\columnwidth]{figures/GandGPi.eps}
\includegraphics[width=1\columnwidth]{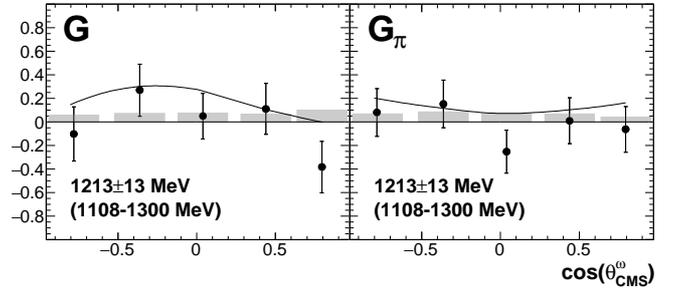}
  \end{center}
  \caption{\footnotesize %
    Polarisation observables, $G$ and $G_\pi$ versus $\cos(\theta^\omega_{CMS})$ at an average beam energy of $1213\pm 13$~MeV (over a range of 1108-1300 MeV).  Systematic errors are on the abscissa.  The solid line is the result of the
    Bonn-Gatchina PWA when including this data (see text for details).  The data are tabulated in Ref.~\cite{EBE}.
  }
  \label{fig:resultsG}
\end{figure}

Data for the observables, $G$ and $G_\pi$ are shown in Fig.~\ref{fig:resultsG}.  Both observables yield small values in the given mass range, compatile with zero. The BnGa fit, described below, reproduces the measured values.

A partial wave analysis was performed in the framework of the Bonn-Gatchina PWA. A large body of data on pion and photo-induced reactions was included which defines masses, widths, and coupling constants of nucleon and $\Delta$ resonances.  New data on $\omega$ photoproduction, which includes
 differential cross sections, density matrix elements \cite{Wilson:2015tbd}, the beam asymmetry 
\cite{AJ,FK}, and the present measurement of the observables $E$, $G$, and $G_\pi$ were also included. The fit returned a $\chi^2=2300$ for the 2072 data points on $\omega$ photoproduction.
The total cross
section receives a large contribution from Pomeron exchange. This contribution rises rapidly from threshold and makes up about $50$\% of the total cross section at 2\,GeV. 
Pion exchange has only a small contribution to the cross section.  Depending on the form factor used, the contribution is between 5-10\% when fitted as a free parameter, however it can be forced to 20\% without deterioration to the description of the data \cite{andrey}.  In addition, the production of baryon resonances is found to be important. Below 1.9\,GeV, the $J^P=3/2^+$ partial wave provides the strongest contribution. If this partial wave is not included in the fit, $\chi^2$ increases by $512$ units.  A $J^P=5/2^+$ partial wave is found which is also required to describe the data reported in \cite{WI1,WI2}; solutions without this contribution are worse in $\chi^2$ by 460 units.  The contributions from the $J^P=3/2^-$ partial wave improve the fit by 331 units in $\chi^2$. 
Within the framework of the PWA, $u$-channel contributions were found to be weak.
A full account of the partial wave analysis, the nucleon resonances contributing to $\gamma p\to\omega p$, and $N^*\to \omega N$ branching ratios will be given elsewhere \cite{Denissenko:2015tbd}.

It is interesting to note in Fig.~\ref{fig:resultsEenergy}, the structure in $E$ at a beam energy of approximately 1650 MeV, where there is evidence of a change of sign from negative to positive at $\cos(\theta_{CMS}^\omega) = +0.125$ and a peak like structure at $\cos(\theta_{CMS}^\omega) = -0.375$ and $-0.125$.  This is close to the $K^*$ threshold, where a cusp-like structure was observed in $K^0\Sigma^+$ photoproduction \cite{ewald:2012, ewald:2014}.  It was speculated that the structure in the $K^0\Sigma^+$ channel may be related to $K^*$ $t$-channel mechanisms, or dynamically $K^*$-hyperon quasi bound states \cite{Ramos13}.

\section{Summary and outlook}
\label{labelSA}

The first measurements of the double polarisation observables $E$, $G$, and $G_\pi$ for $\gamma p \to p \omega$ have been reported. The beam-target-helicity asymmetry $E$ was measured from threshold to a photon energy of 2300~MeV, and $G$ and $G_\pi$ were measured at a single bin in photon energy at $1108<E_\gamma<1300$\,MeV. The results clearly show that
$s$-channel contributions, in addition to the expected $t$-channel contributions, have significant
importance in $\omega$ photoproduction close to threshold.

A fit to the data within the framework of the Bonn-Gatchina partial wave analysis requires significant contributions of the partial waves with $J^P=3/2^+, 5/2^+$, and $3/2^-$ to $\omega$ photoproduction.

A possibility to improve statistics in the $\omega$ channel is to exploit
the mixed charged decay ($\omega\rightarrow\pi^{+}\pi^{-}\pi^{0}$) with a
branching ratio of 89.2~\% \cite{PDG}. This cannot be done within the present
CBELSA/TAPS setup but will instead be pursued with the new BGO-OD experiment \cite{bantes13, HS10-BGOOD, HS} 
at ELSA.  The BGO-OD experiment will also be used to analyse other vector
meson channels (for example $\phi$ and $K^{*}$ production) off the proton and neutron, in order
to study $t$-channel exchange processes and the contributions from nucleon resonances in greater detail.

\section{Acknowledgments}
We thank the technical staff of ELSA and the participating institutions for
their invaluable contributions to the success of the experiment. We
acknowledge support from the Deutsche Forschungsgemeinschaft (SFB/TR16) and
Schweizerischer Nationalfonds.

%% The Appendices part is started with the command \appendix;
%% appendix sections are then done as normal sections
%% \appendix

%% \section{}
%% \label{}

%% References
%%
%% Following citation commands can be used in the body text:
%% Usage of \cite is as follows:
%%   \cite{key}          ==>>  [#]
%%   \cite[chap. 2]{key} ==>>  [#, chap. 2]
%%   \citet{key}         ==>>  Author [#]

%% References with bibTeX database:

\bibliographystyle{model1a-num-names}
%\bibliography{<your-bib-database>}
%\bibliography{omega_paper}

%% Authors are advised to submit their bibtex database files. They are
%% requested to list a bibtex style file in the manuscript if they do
%% not want to use model1a-num-names.bst.

%% References without bibTeX database:

\end{document}